\documentclass[runningheads]{llncs}
\usepackage[T1]{fontenc}
\usepackage[autoload=true, charsperline=60]{jlcode}

\usepackage{float}

\usepackage[]{todonotes}   % add 'disable' to hide notes

\usepackage{graphicx}
\usepackage{hyperref}

\newif\ifanonymous
\anonymousfalse

\begin{document}
\title{Reactive Environments for Active Inference Agents with RxEnvironments.jl}

\ifanonymous

\authorrunning{Anonymous Authors}
\author{Anonymous Authors
\authorrunning{Anonymous Authors}
\institute{Undisclosed Affiliation}
% \email{\{w.w.l.nuijten, bert.de.vries\}@tue.nl}\\
}

\else
\authorrunning{W. W. L. Nuijten \& B. de Vries}
\author{Wouter W.L. Nuijten\inst{1}
\orcidID{0009-0007-0689-9300}
\and 
Bert de Vries\inst{1,2}
\orcidID{0000-0003-0839-174X} 
\authorrunning{W. W. L. Nuijten \& B. de Vries}
\institute{Eindhoven University of Technology, 5612 AP Eindhoven, the Netherlands \and  GN Hearing, 5612 AB Eindhoven, The Netherlands}
\email{\{w.w.l.nuijten, bert.de.vries\}@tue.nl}\\
}
\fi
\maketitle              % typeset the header of the contribution

\begin{abstract}
Active Inference is a framework that emphasizes the interaction between agents and their environment. While the framework has seen significant advancements in the development of agents, the environmental models are often borrowed from reinforcement learning problems, which may not fully capture the complexity of multi-agent interactions or allow complex, conditional communication. This paper introduces Reactive Environments, a comprehensive paradigm that facilitates complex multi-agent communication. In this paradigm, both agents and environments are defined as entities encapsulated by boundaries with interfaces. This setup facilitates a robust framework for communication in nonequilibrium-Steady-State systems, allowing for complex interactions and information exchange. We present a Julia package \jlinl{RxEnvironments.jl}, which is a specific implementation of Reactive Environments, where we utilize a Reactive Programming style for efficient implementation. The flexibility of this paradigm is demonstrated through its application to several complex, multi-agent environments. These case studies highlight the potential of Reactive Environments in modeling sophisticated systems of interacting agents.

\end{abstract}
%
%\tableofcontents
% \bdv{make header capitalization consistent. I added a TOC for clarity of structure. You can delete it in final version. }

\section{Introduction} \label{sec:introduction}
The Free Energy Principle (FEP) \cite{friston_free-energy_2010} distinguishes itself from other theories of self-organization by taking an interaction-centric perspective. Active inference (AIF) is an implication of the Free Energy Principle that extends the FEP to control and decision-making in self-organizing natural systems.

In this framework, agents possess an internal generative model for predicting observations from an unknown external process. The model updates its internal (perceptive) and active (control) states to minimize prediction errors. This unifying principle has profound implications for understanding how agents perceive and act within complex environments.

AIF posits that agents actively seek to minimize their free energy (a measure related to surprise or prediction error) by updating their beliefs about the environment and selecting actions that align with these beliefs \cite{friston_action_2010}. This formulation bridges the gap between theoretical principles and practical implementations of FEP in agent-environment interactions.

To simulate a synthetic AIF agent, researchers need the ability to control interactions between agents and their environment in practical scenarios. For example, a significant theory from AIF is that the human brain learns from the proprioceptive feedback it receives from muscles \cite{adams_predictions_2013}. Since proprioceptive and exteroceptive sensory channels do not necessarily run at the same time-frequency rates, researchers need fine-grained control over the communication protocol between the agent and the environment. In other settings, one could be interested in having multiple agents share the same world, allowing communication between agents \cite{friston_federated_2024}. Current solutions from the reinforcement learning or control theory community, such as Gymnasium \cite{towers_gymnasium_2024} do not give end-users these controls over details of the environment, instead focusing on implementing a single agent-environment interaction through a transition function. The imperative programming style used in these frameworks limits the communication between agents and environments with a predefined time step, observation frequency, and action frequency.

This paper introduces Reactive Environments, which adopt a reactive programming approach to environment design. In contrast to their imperative counterparts, Reactive Environments are not limited by strict communication constraints and natively allow multi-sensor, multimodal interaction between agent and environment. We will discuss how a reactive programming strategy addresses the flaws of current frameworks and introduce \emph{\href{https://www.github.com/biaslab/rxenvironments.jl}{RxEnvironments.jl}}, a specific implementation of Reactive Environments in the Julia language \cite{bezanson_julia_2015}. We will show how implementing complex real-world environments with fine-grained control over an agent's observations is streamlined in RxEnvironments.jl. The main features of RxEnvironments.jl are:
\begin{itemize}
    \item Detailed control over observations. Different sensory channels can execute at different frequencies or can be triggered only when specific actions are taken, allowing for complex interactions.
    \item Native support for multi-agent environments: multiple instances of the same agent type can be spawned in the same environment without additional code.
    \item Reactivity: By employing a reactive programming style, we ensure that environments will emit observations when prompted, and will idle when no computation is necessary.
    \item Support for multi-entity complex environments where the agent-environment framework does not suffice.
\end{itemize}
With RxEnvironments, we hope to contribute to standardizing the creation and simulation of Active Inference agents, allowing researchers to share their environments and potentially creating standardized benchmarks in the future.

The main contributions of this paper are as follows: 
\begin{itemize}
    \item We define the Reactive Environment concept in Section~\ref{sec:reactive-env-def}. 
    \item In Section~\ref{sec:rxenvironments-method}, we introduce RxEnvironments as a package to create environments for Active Inference agents.
    \item In Section~\ref{sec:case_study}, we demonstrate how to create complex environments with unique needs.
\end{itemize}
\section{Related Work} \label{sec:related_work}

In reinforcement learning, the creation and sharing of control environments has mainly been standardized with the introduction of Gymnasium \cite{towers_gymnasium_2024}. Users can use the step function in Gym to define a transition function, and Gym will handle the environmental simulation. A similar alternative in Python, based on the MuJoCo physics engine, is Deepmind Control Suite \cite{tassa_deepmind_2018}. The equivalent alternative in the Julia programming language would be ReinforcementLearning.jl \cite{tian_reinforcementlearning_2020}. These packages export high-level interfaces to the environments they describe, alleviating the user's burden of timekeeping. Although these packages are designed explicitly for reinforcement learning, which involves computing a reward metric at every state, they can also be used for Active Inference as they describe general control environments \cite{ueltzhoffer_deep_2018,van_de_maele_integrating_2023}. Although the realization of environments is also part of popular packages such as PyMDP \cite{heins_pymdp_2022} and the SPM-DEM toolbox \cite{friston_dem_2008}, these packages have their primary focus on agent creation. As a result, we do not present RxEnvironments as a substitute but rather as a comprehensive framework that agent-centric packages can use.

In general, we observe that there is no standardized way of defining environments for Active Inference agents. While some implementations use Gymnasium \cite{ueltzhoffer_deep_2018,van_de_maele_integrating_2023,safa_active_2023}, others use specialized toolboxes for implementing Active Inference agents for their environment simulation \cite{esaki_dynamical_2024,friston_federated_2024}. With RxEnvironments, we aim to unite all use cases for environments in Active Inference in a robust and comprehensive package.

Reactive Programming (RP) has wide applications in various domains and is similar to the Actor Model \cite{hewitt_session_1973}. RP does not assume anything about the data generation process, allowing computation both on static datasets and real-time asynchronous sensor observations. In Reactive Programming, it is necessary to define how the system should react to changes in data or events rather than explicitly programming sequences of steps. This approach is similar to an in situ control system that can gather data through its sensors asynchronously and respond accordingly to incoming stimuli. 
\section{Methods} \label{sec:method}

\subsection{A Model for Interaction in Active Inference}
\begin{figure}[ht]
    \centering
    \includegraphics[width=\textwidth]{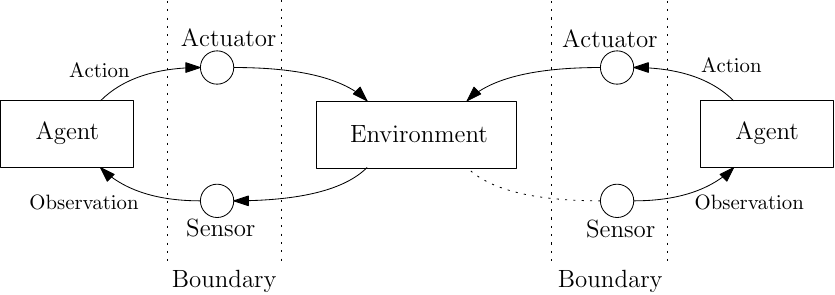}
    \caption{General communication protocol in an Active Inference environment containing two agents. The terms "Actuator" and "Sensor" are used from the agents' points of view. We see that both agents have a boundary with actuators and sensors with which they interact with the environment.}
    \label{fig:method-markov-blankets}
\end{figure}

Self-organizing agents maintain their existence by creating a boundary that separates their internal states from the external states of their environment \cite{hesp_multi-scale_2019,varela_autopoiesis_1974,palacios_markov_2020,kirchhoff_markov_2018}. An agent can only affect its environment through its actuators, while its internal states are influenced only by stimuli received through its sensors. Therefore, an agent has a set of actuator and sensor interfaces that it uses to communicate with its environment. We will refer to this collection of all actuators and sensors of an agent as its boundary. A schematic of this interaction model can be seen in Figure~\ref{fig:method-markov-blankets}.

Here, the duality between the agent and environment is notable; the actions emitted by the agent are perceived as observations for the environment, and vice versa. This dual separation of internal and external states has prompted an overarching term for agents and environments, which we call an \textbf{Entity}. Note that Entities are separate from Fristonian "things" \cite{friston_free_2019} in the sense that an Entity can be any "thing" or an environment, making it a superclass to things. We use the following definition of an Entity:
\begin{definition}{Entity.}
An Entity is a structure with a set of actuators and sensors called a boundary that allows it to communicate with other Entities. 
\end{definition}
An Entity can, but is not obliged to, have an internal state that the sensor interfaces in its boundary can influence. Entities with connected pairs of actuators and sensors are mutually "subscribed." We use this term because any action emitted by an agent prompts a change in the internal state of an environment. Generally, an emitted action by an Entity prompts activity (either a response or an update of an internal state) from the subscribed entities. In Figure \ref{fig:method-markov-blankets}, we see three entities, two agents that are both subscribed to the same environment Entity, with the boundaries of both agents expanded. 

The notion of a Markov Blanket is prevalent in Active Inference literature \cite{palacios_markov_2020,kaufmann_active_2021}, and it denotes the statistical partition between a system's internal and external states. It is worth noting that the concept of a boundary, as formulated in this section, coincides with the notion of a Markov Blanket used in Active Inference. Therefore, modeling communication as an interaction between different Entities through their boundaries is an adequate implementation of the communication of a probabilistic model with a Markov Blanket and its environment.

\subsection{Reactive Environments} \label{sec:reactive-env-def}
\begin{figure}[t]
    \centering
    \includegraphics{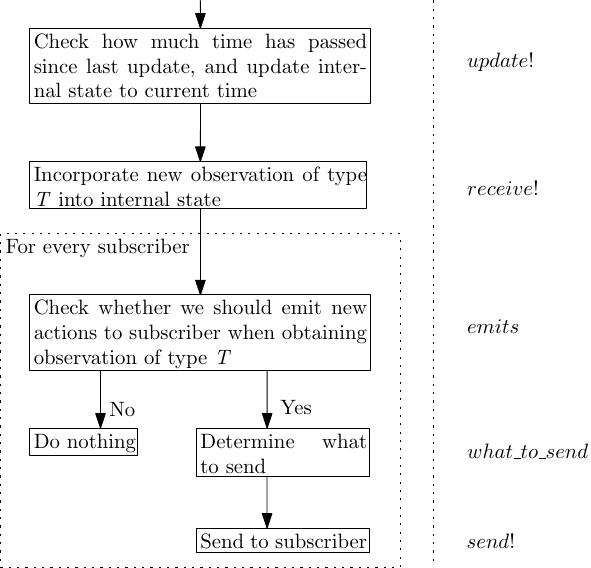}
    \caption{Internal Entity logic is applied when an observation is received. On the left, we outline the steps an Entity should follow when processing an observation. On the right, we specify the \jlinl{RxEnvironments} functions that users can create to customize this behavior.}
    \label{fig:observation_flowchart}
\end{figure}

Communication between Entities flows through their respective boundaries. Any Entity can send data through its actuator interface to its subscribers at any point, prompting activity in subscribed entities. Building upon the discussion in Section~\ref{sec:related_work} regarding the Reactive Programming paradigm, we extend the concept to environments in agent-based systems. Entities should process sensory data seamlessly and respond to subscribers. Just as programmers define how systems respond to impulses, we aim to define how entities react to impulses exerted by the entities to which they are subscribed. To this extent, we define a Reactive Environment:

\begin{definition}{Reactive Environments.}
A Reactive Environment is a pair $\mathcal{E} = (\mathcal{A}, \mathcal{S})$ where $\mathcal{A}$ is a set of Entities and $\mathcal{S}$ is a mutable set of subscriptions, where every $s \in \mathcal{S}$ is a pair $(\mathcal{A}_1, \mathcal{A}_2)$, $\mathcal{A}_2, \mathcal{A}_2 \in \mathcal{A}$. Each Entity responds reactively to sensory impulses received from any of its subscribers through its sensors and is able to emit data to any subscriber through its actuators. 
\end{definition}
For Entities in a Reactive Environment, we provide a set of desiderata that enable the design of complex communication networks within the Reactive Environment framework. 
\begin{itemize}
    \item Entities should be able to update their internal state in response to received impulses. The update should be based either on the impulse emitter or the type of observation. For example, an agent should update its internal state differently when receiving an audio signal versus a video signal.
    \item An Entity should be able to determine whether or not to transmit any received impulse to its subscribers. For example, suppose an agent sends a proprioceptive signal, and the Entity representing the environment receives it. In that case, the environment can match it with an observation, but it does not have to transmit it to its other subscribers.
    \item At any given time, an Entity should be able to send a signal to its subscribers, such as a video camera emitting a 60Hz signal.
    \item An Entity should be able to send different signals to different subscribers when emitting. For example, an environment Entity can send different observations to different agent entities based on their relative position in the environment.
\end{itemize}
In Figure \ref{fig:observation_flowchart}, we see a flow chart of the logic we want every Entity to go through whenever they get an observation. This logic can also be triggered regularly to mimic a sensor continuously providing data at a fixed rate. An algorithm implementing the logic in this flowchart allows for all the behaviors listed above.

\subsection{RxEnvironments.jl: a Particular Implementation of Reactive Environments} \label{sec:rxenvironments-method}
In this section, we introduce \jlinl{RxEnvironments.jl}\footnote{https://github.com/biaslab/RxEnvironments.jl}, a package in the open-source Julia \cite{bezanson_julia_2015} language that implements the communication protocol and the desiderata described in Section~\ref{sec:reactive-env-def}. In RxEnvironments, we take an Entity-centric standpoint and implement all communication logic on the Entity level. This means that entities representing agents and environments have no constraints on their communication, allowing agent-agent subscriptions or multi-agent environments natively. The reactive programming features of RxEnvironments are based on the \emph{Rocket.jl} Reactive Programming library \cite{bagaev_rocketjl_2020}. In Figure~\ref{fig:observation_flowchart} we see the flow chart that describes the logic entities go through when processing observations, on the right we see the corresponding functions in \jlinl{RxEnvironments}. In this section, we will go through these functions in more detail.

\subsubsection{Revising the transition function}

In popular environment design frameworks, such as Gymnasium \cite{towers_gymnasium_2024}, the transition function for an environment takes the action emitted by an agent and the previous state. It produces a new state and an observation for the agent. In our Entity-centric approach, this modeling assumption is restrictive for various reasons:
\begin{itemize}
    \item In a scenario where entities can have multiple subscribers and be subscribed to multiple entities, it is not desirable to trigger the transition function for every received stimulus. To illustrate, consider a multi-agent environment where all actions emitted by each agent should be gathered before updating and advancing the environment time. This ensures that the transition function is not invoked too frequently.
    \item The transition function cannot have a constant time interval. It should be a function of the elapsed time since the last update to account for stimuli being received at any time.
\end{itemize}

For these reasons, we split the transition update function into two different distinct parts:
\begin{enumerate}
    \item  The function \jlinl{receive!} updates the internal state of the receiving Entity by incorporating the stimulus received, given the emitting and receiving entities and the stimulus as inputs. 
    \item The function \jlinl{update!} takes an Entity and the time elapsed since the last update of this Entity and updates the internal state of the Entity to reflect that additional time has elapsed.
\end{enumerate}
By splitting these two functions, we can natively support multi-agent environments, and we can simulate the behavior of the internal state of an Entity in the absence of stimuli. A concrete example would be that multiple agents could still observe a ball bouncing in an environment without explicitly abstaining from action at every time step since the dynamics of the bouncing ball is described in the \jlinl{update!} function, which will continue to be called even in the absence of stimuli.

\subsubsection{Control over emission logic}

In Section~\ref{sec:reactive-env-def}, we have emphasized the importance of controlling the emission logic when an Entity receives a stimulus. Therefore, we expose two functions that govern emission logic for entities in RxEnvironments:
\begin{enumerate}
    \item The \jlinl{emits} function operates by taking in the receiving Entity, the stimulus data received by the Entity, and any subscribers of the Entity. The function then returns a value determining whether the Entity should emit the stimulus to a particular subscriber. This function is called for all subscribers.
    \item The function \jlinl{what_to_send} takes the receiving Entity, the stimulus data that the Entity receives, and any subscriber of the Entity. This function is called when \jlinl{emits} returns true and determines which stimulus will be sent to that subscriber.
\end{enumerate}
In this way, designers of environments have full control over when and how their entities should emit. In the next section, we will demonstrate the versatility of this framework by designing several complex environments.

\subsubsection{Internal triggers for emission}

Of course, not all emissions in Entity interactions are triggered by external impulses. Sensors, for example, usually provide data at a steady frequency. Therefore, we expose an interface to emit observations to all subscribers at regular intervals. By creating entities for different sensors and attaching them to an overarching Entity, we can create complex systems that combine data from different sensors at different observation frequencies.

\subsubsection{Replicating classical reinforcement learning environments}

In classical reinforcement learning environments, the environment is often seen as a passive recipient of actions from the agent, responding to actions with observations in the next time step. A different way to view environments is to consider them as reactive environments. In this case, the Entity representing the environment waits until all subscribed entities have emitted before calling the state transition function at a predetermined time interval. This approach transforms classical reinforcement learning environments into specific instances of reactive environments, making environmental simulation more flexible and generalizable.
\section{Case Studies} \label{sec:case_study}
In this section, we will implement several increasingly complex environments. In every environment, we employ a specific strategy from Reactive Environments that cannot be replicated in popular environment creation packages. \footnote{
\ifanonymous
The implementation of the environments discussed in this section will be provided in the final version to guarantee anonymity throughout the review process.
\else
The implementation of the environments discussed in this section can be found at \url{https://github.com/wouterwln/RxEnvironments-Examples}
\fi
}

\subsection{Mountain Car}
\begin{figure}[t]
    \centering
    \includegraphics[width=\textwidth]{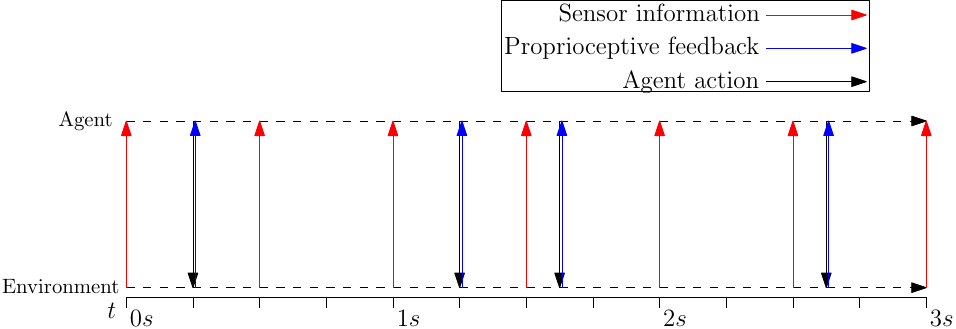}
    \caption{Overview of interactions in the Mountain Car environment over time. The environment emits sensor information at a regular interval (2 Hz in this example), and whenever the agent emits an action, the environment instantaneously responds with proprioceptive feedback to the agent.}
    \label{fig:mountaincar-interaction}
\end{figure}
In this section, we implement a classic environment in reinforcement learning, the Mountain Car environment \cite{ueltzhoffer_deep_2018}, with a slight difference: whenever an agent emits an action (e.g., setting the throttle of the engine), we match this with an observation from the environment that contains the engine force applied. For example, if the agent decides to apply $300\%$ of its engine force, the environment will reply that only $100\%$ is applied. Presenting observations in this way aligns with \cite{adams_predictions_2013}. To realize this, we design our environment to trigger different implementations of \jlinl{what_to_send} based on the input stimulus: Whenever a throttle action is received, we return the throttle action that is applied to the environment, and we let the environment emit on a regular frequency of 2 Hz to replicate a sensor that measures the position and velocity of the mountain car. In Figure \ref{fig:mountaincar-interaction}, we see a schematic overview of the interactions between the agent and the environment. We see that we have two distinct types of observations from the agent, sensory and proprioceptive feedback, and the logic for obtaining both is different.

\subsection{Football Environment}

\begin{figure}[t]
    \centering
    \includegraphics[width=0.9\textwidth]{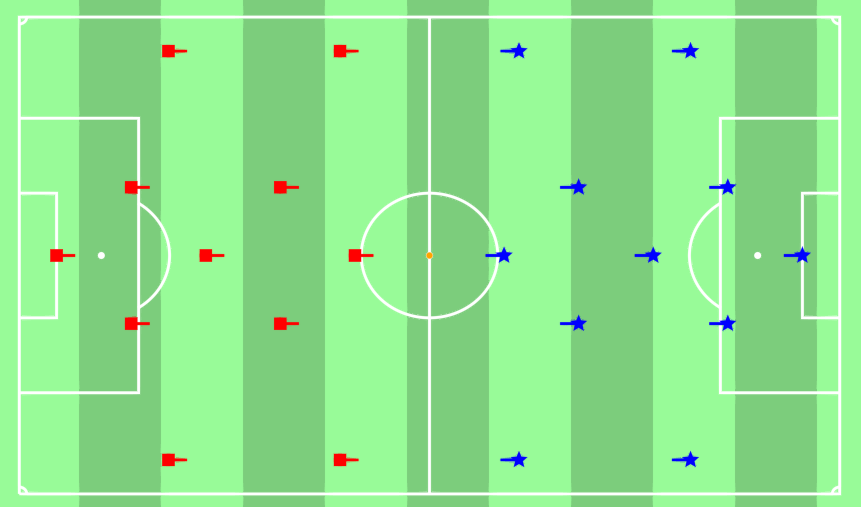}
    \caption{Plot of the setup of our football environment, showing the pitch and the 22 players. The ball is positioned on the center spot. An animation of this environment where we send random run commands to players can be found \href{https://youtu.be/24ZSPcVDOqc}{here}.}
    \label{fig:case-study-football-env}
\end{figure}

Next, we describe the implementation of a simulation environment of a football match. Football is the most popular sport in the world and involves two teams of 11 players who handle a ball with their feet to score goals. A football game is a complex multi-agent game where the entity representing the environment has to handle inputs from all 22 agents to let the game run smoothly. Additionally, since we are in a noncooperative game between two teams, players can emit signals (shout) to their teammates, so we also have an agent-to-agent communication channel. Since all players have their position and orientation on the pitch, their field of vision and the observations they receive are also different for each agent.

We model this environment with a single Entity representing the state of the world and 22 Entities representing the individual players. The world contains the ball and the references to all 22 player bodies, so collisions and on-ball actions can be resolved. All player Entities are subscribed to the world Entity but are not subscribed to each other. We do not explicitly model agent-to-agent interactions because this would unnecessarily complicate the subscription graph. Instead, a player can choose to emit a signal to all other players, which the world Entity will forward to all other players. In a sense, the world Entity represents the "global" state of the system that keeps track of all physical interactions. At the same time, all player Entities contain their local states and receive observations from the global state. In Figure \ref{fig:case-study-football-env}, we have visualized an example of this environment. This example shows that we can, with the code used to define a single player, create a 22-player environment. In \href{https://youtu.be/24ZSPcVDOqc}{this YouTube video}, we show that we can send commands to all individual players asynchronously. Here, we send the command to run in a random direction to a random player every 0.1 seconds.

Due to the intricate dynamics of the football game and its non-trivial set of rules, we have decided only to model running and on-ball actions. We aim to demonstrate the multi-agent nature of Reactive Environments rather than create a comprehensive football environment.

\subsection{Hearing Aid Environment}
\begin{figure}[t]
    \centering
    \includegraphics[width=\textwidth]{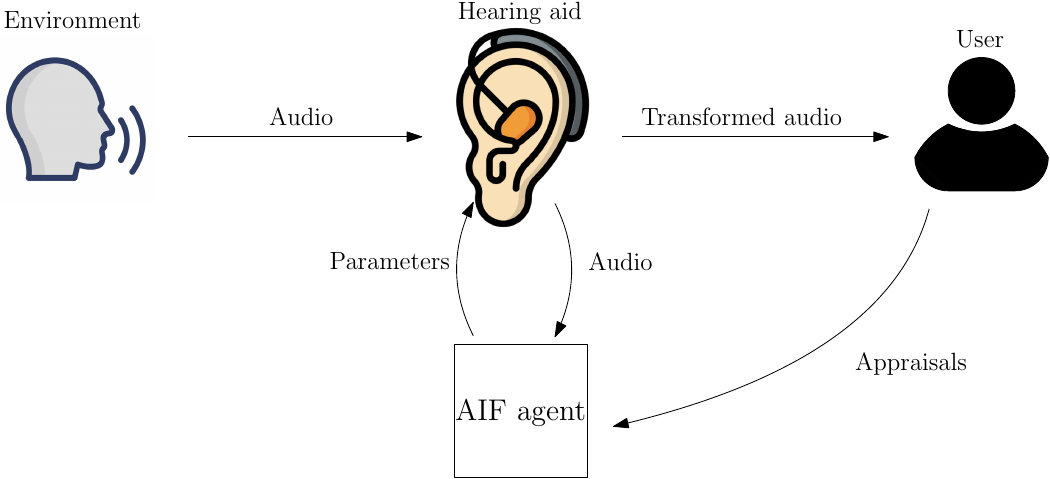}
    \caption{Schematic of the subscriptions in the hearing aid environment.}
    \label{fig:case-study-hearing-aid}
\end{figure}

Hearing aids often feature advanced acoustic noise reduction algorithms. In recent years, we have seen the rise of active inference-based agents that parameterize hearing aid noise reduction algorithms \cite{podusenko_aida_2022}. Since a hearing aid has very limited computing power and battery capacity, sometimes part of the agent's needed computations must be performed on a separate wearable device, e.g., the patient's phone. This configuration leads to a unique multi-entity system where the hearing aid is continually communicating with three different entities: (1) the outside world, which emits acoustic signals; (2) the user (hearing aid patient), who receives the hearing aid output signal (and can potentially emit feedback to the hearing aid about the perceived performance of the hearing aid); and (3) with the intelligent agent at the user's phone. Figure \ref{fig:case-study-hearing-aid} shows a schematic overview.

Thus, we obtain a complex entity interaction, where the hearing aid can obtain stimuli from all 3 subscribed entities and should process the data accordingly: an acoustic signal from the outside world should be processed and emitted to both the user and the agent; a new proposal for parameter settings, i.e., the agent's actions, should be incorporated into the signal processing algorithm; and user appraisals should also be forwarded to the agent that will use these appraisals to update its future parameter proposals.

In short, we have a complex multi-entity interaction where every entity should handle different stimuli in different ways. We have to control which signal to emit and when to emit the signal. Note that the hearing aid should not send a signal to the user when receiving a new set of parameters but only when the sound from the outside environment is registered. In a Reactive Environment, all interactions are well-defined, and we can observe the signal the user hears while also designing an agent that would take the place of the agent in our setting.
\section{Discussion}
While Reactive Environments generalize environments for learning agents, the design of agents that can interact with a Reactive Environment should still be investigated. This is because, in our framework, there are fewer constraints on the communication between an agent and its surroundings. Traditionally, our agent receives 1 observation per predetermined timestep and can process this observation accordingly. In a Reactive Environment, agents can receive (or not receive) data at any point in time, necessitating an internal clock for the agents themselves. Although relieving this constraint on communication allows for interesting experiments (we can disable a sensor for a particular agent to simulate the sensor breaking down and investigate how our agent handles this loss of data), the authors are not aware of any implementation of agents that are built for this level of flexibility. An interesting avenue is Reactive Message Passing on a Factor Graph \cite{loeliger_factor_2007,bagaev_reactive_2021}, which employs the same reactive programming strategy to Bayesian inference that we have taken in this paper to environment design.

\section{Conclusions}
In this paper, we presented the concept of a Reactive Environment and a particular implementation, \jlinl{RxEnvironments.jl}. We showed that environments defined in the classical reinforcement learning literature can be written as particular cases of Reactive Environments, and we showed that we can model more complex interactions within this paradigm. In particular, we showed that with Reactive Environments we are able to model the complex communications between agents and environments necessary to realize Active Inference simulations. In our case studies, we showed that our framework can be used to define a multitude of different environments, demonstrating the expressive power of the framework. Furthermore, we have presented \jlinl{RxEnvironments.jl}, a particular implementation of Reactive Environments. Extensions of this work might investigate the classes of agents that handle the communication protocol employed by Reactive Agents to simulate how agents would operate in the field.

\section*{Acknowledgements}
This publication is part of the project "ROBUST: Trustworthy AI-based Systems for Sustainable Growth" with project number KICH3.LTP.20.006, which is (partly) financed by the Dutch Research Council (NWO), GN Hearing, and the Dutch Ministry of Economic Affairs and Climate Policy (EZK) under the program LTP KIC 2020-2023.

The authors thank Thijs van de Laar, Magnus Koudahl, and Tim Nisslbeck for their insightful discussions during the project's execution.

\bibliographystyle{splncs04}
\bibliography{references}

\appendix
\section{Appendix: Creating a simple environment}
In this code example, we will demonstrate the creation of a simple environment in \jlinl{RxEnvironments}, demonstrating that the additional boilerplate code needed to write an imperative environment as a Reactive Environment is minimal. We will implement the Bayesian Thermostat example, which is also showcased in the \href{https://biaslab.github.io/RxEnvironments.jl/stable/lib/getting_started/}{RxEnvironments documentation}.
\subsection{Defining the environment}
The Bayesian Thermostat environment is a very simple environment that monitors the temperature in a room. The temperature can fluctuate between a minimal and a maximal temperature, and an agent can influence this temperature by adding or subtracting heat from the room. Furthermore, the environment cools down over time.
\subsection{Environment boilerplate}
In this section we will write all boilerplate code necessary to run the environment. We start by defining the structures needed to store the temperature and environment properties and expose helper functions that change this temperature, namely the \jlinl{add_temperature!} function. 
\begin{jllisting}
using Distributions

# Empty agent, could contain states as well
struct ThermostatAgent end

mutable struct BayesianThermostat{T}
    temperature::T
    min_temp::T
    max_temp::T
end

# Helper functions
temperature(env::BayesianThermostat) = env.temperature
min_temp(env::BayesianThermostat) = env.min_temp
max_temp(env::BayesianThermostat) = env.max_temp
noise(env::BayesianThermostat) = Normal(0.0, 0.1)
set_temperature!(env::BayesianThermostat, temp::Real) = env.temperature = temp
function add_temperature!(env::BayesianThermostat, diff::Real)
    env.temperature += diff
    if temperature(env) < min_temp(env)
        set_temperature!(env, min_temp(env))
    elseif temperature(env) > max_temp(env)
        set_temperature!(env, max_temp(env))
    end
end
\end{jllisting}
\subsection{RxEnvironments specific code}
In this section, we will implement the \jlinl{RxEnvironments}-specific code. We have a very simple interaction scheme: When the agent emits an action, we want this to be incorporated into the environment state, but we only want the environment to emit observations on a fixed frequency, and not present an observation whenever the agent chooses to change the environment. Therefore, we implement the \jlinl{receive!}, \jlinl{update!}, \jlinl{emits} and \jlinl{what_to_send} functions for the environment:
\begin{jllisting}
# When the environment receives an action from the agent, we shouldn't emit back to the agent
RxEnvironments.emits(::BayesianThermostat, ::ThermostatAgent, ::Real) = false

# In any other case, we should emit (This line is obsolete since this is the default behavior, but we include it for clarity)
RxEnvironments.emits(::BayesianThermostat, ::ThermostatAgent, any) = true

# When the environment receives an action from the agent, we add the value of the action to the environment temperature.
RxEnvironments.receive!(recipient::BayesianThermostat, emitter::ThermostatAgent, action::Real) = add_temperature!(recipient, action)

# The environment sends a noisy temperature observation to the agent.
RxEnvironments.what_to_send(recipient::ThermostatAgent, emitter::BayesianThermostat) = temperature(emitter) + rand(noise(emitter))

# The environment cools down over time.
RxEnvironments.update!(env::BayesianThermostat, elapsed_time)= add_temperature!(env, -0.1 * elapsed_time)
\end{jllisting}
\subsection{Invoking the environment}
We now have all the code necessary to kickstart our environment:
\begin{jllisting}
environment = RxEnvironment(BayesianThermostat(0.0, -10.0, 10.0); emit_every_ms = 1000)
agent = add!(environment, ThermostatAgent())
\end{jllisting}
Now, your environment will be running, and \jlinl{agent} will receive a noisy observation from the environment every second.
\end{document}